\documentclass[prl, aps, notitlepage, superscriptaddress, showpacs, a4paper,twocolumn,10pt]{revtex4-1} 

\usepackage{graphicx}  
\usepackage{bm}        
\usepackage{amssymb}   
\usepackage[latin1]{inputenc}
\usepackage[T1]{fontenc}
\usepackage[english]{babel}
\usepackage{subfigure}
\usepackage{epsfig}
\usepackage{verbatim}
\usepackage{setspace}
\usepackage{placeins}
\usepackage{amsmath}
\usepackage{mathrsfs}
\usepackage{slashed}
\usepackage[usenames, dvipsnames]{color}
\usepackage{graphics}
\usepackage[margin=1in,footskip=0.25in]{geometry} 
\usepackage[bookmarks=false,pdfborder={0 0 0}]{hyperref}
\usepackage[mediumspace,mediumqspace,squaren]{SIunits}
\usepackage{amsfonts}
\hypersetup{
colorlinks=true,
linkcolor=blue,
citecolor=blue}

\newcommand\abs[1]{\left|#1\right|}

\begin{document}

\title{Injection of orbital angular momentum and storage of quantized vortices in polariton superfluids}
\author{T.~Boulier}
\affiliation{Laboratoire Kastler Brossel, UPMC-Sorbonne Universit\'{e}s, CNRS, ENS-PSL Research University, Coll\`ege de France, 4, place Jussieu Case 74, F-75005 Paris, France.}
\author{E.~Cancellieri}
\affiliation{Department of Physics and Astronomy, University of Sheffield,\\
Hicks Building, Hounsfield Road, Sheffield, S3 7RH, England.}
\author{N.D.~Sangouard}
\affiliation{Laboratoire Kastler Brossel, UPMC-Sorbonne Universit\'{e}s, CNRS, ENS-PSL Research University, Coll\`ege de France, 4, place Jussieu Case 74, F-75005 Paris, France.}
\author{Q.~Glorieux}
\affiliation{Laboratoire Kastler Brossel, UPMC-Sorbonne Universit\'{e}s, CNRS, ENS-PSL Research University, Coll\`ege de France, 4, place Jussieu Case 74, F-75005 Paris, France.}
\author{A.V.~Kavokin}
\affiliation{School of Physics and Astronomy, University of Southampton,
Highfield, Southampton, SO17 1BJ, United Kingdom.}
\affiliation{CNR-SPIN, Viale del Politecnico 1, Rome I-00133, Italy.}
\author{D.M.~Whittaker}
\affiliation{Department of Physics and Astronomy, University of Sheffield,\\
Hicks Building, Hounsfield Road, Sheffield, S3 7RH, England.}
\author{E.~Giacobino}
\affiliation{Laboratoire Kastler Brossel, UPMC-Sorbonne Universit\'{e}s, CNRS, ENS-PSL Research University, Coll\`ege de France, 4, place Jussieu Case 74, F-75005 Paris, France.}
\author{A.~Bramati}
\email{bramati@lkb.umpc.fr}
\affiliation{Laboratoire Kastler Brossel, UPMC-Sorbonne Universit\'{e}s, CNRS, ENS-PSL Research University, Coll\`ege de France, 4, place Jussieu Case 74, F-75005 Paris, France.}

\date{\today}

\begin{abstract}
We report the experimental investigation and theoretical modeling of a rotating polariton superfluid relying on an innovative method for the injection of angular momentum.
This novel, multi-pump injection method uses four coherent lasers arranged in a square, resonantly creating four polariton populations propagating inwards. 
The control available over the direction of propagation of the superflows allows injecting a controllable non-quantized amount of optical angular momentum. 
When the density at the center is low enough to neglect polariton-polariton interactions, optical singularities, associated to an interference pattern, are visible in the phase. In the superfluid regime resulting from the strong nonlinear polariton-polariton interaction, the interference pattern disappears and only vortices with the same sign are persisting in the system. Remarkably the number of vortices inside the superfluid region can be controlled by controlling the angular momentum injected by the pumps.
\end{abstract}

\maketitle

\paragraph{Introduction.}
\addcontentsline{toc}{section}{Introduction}
In planar semiconductor microcavities, the strong coupling between light (photons) and matter (excitons)~\cite{Weisbuch92} gives rise to exciton-polaritons, with specific properties such as a low effective mass, inherited from their photonic component, and strong nonlinear interactions due to their excitonic part.
These quasi-particles offer a great opportunity to revisit in solid state materials the interaction between light and matter, first explored {in atomic physics}.
{Polaritonic systems} are easily controllable by optical techniques and, due to their finite lifetimes, are ideal systems for studying out-of-equilibrium phenomena \cite{RevModPhys.82.1489,RevModPhys.85.299}.
In analogy with the atomic case \cite{RevModPhys.71.463,PhysRevLett.86.4447}, the superfluid behavior of polariton quantum fluids has been theoretically predicted \cite{PhysRevLett.93.166401} and experimentally confirmed~\cite{Amo09b, Amo09a, sanvitto10}.

Quantized vortices are topological excitations characterized by the vanishing of the field density at a given point (the vortex core) and the quantized winding of the field phase from~$0$ to~$2\pi$ around it. 
Together with solitons, they have been extensively studied and observed in nonlinear optical systems \cite{desyatnikov2005optical}, superconductors \cite{essmann1967direct}, superfluid $^{4}$He \cite{PhysRev.136.A1194} and more recently in cold atoms \cite{madison2000vortex,denschlag2000generating,khaykovich2002formation}.
Even though vortices have already been theoretically proposed~\cite{Pigeon11} and experimentally observed~\cite{roumpos11, Nardin11, sanvitto11, Lagoudakis08, manni13} in polariton fluids, more detailed studies of vortices and vortex arrays are still needed in order to achieve a better understanding of polariton superfluidity and of vortex dynamics, as well as to achieve the implementation {of quantum technologies}~\cite{zhenghan10, freedman03, nayak08}.

Polariton systems have been shown to reveal a large variety of effects with the formation of stable vortices~\cite{Lagoudakis08,Dall2014} and half vortices \cite{lagoudakis09, flayac10} as well as the formation of single vortex-antivortex (V--AV) pairs \cite{roumpos11, Nardin11, tosi11}, and spin-vortices \cite{spin-vortices15}.
The formation of lattices of vortices and of V--AV pairs has been theoretically predicted for cavity-polaritons \cite{keeling08, Gorbach10, Liew08} and observed in the case of patterns induced by metallic deposition on the surface of the cavity \cite{Kusudo13}.
Such lattices are also observable when the interplay between the excitation shape and the underlying disorder pins the vortices, allowing their detection in time-integrated experiments \cite{manni13}.

\begin{figure}
\begin{center}
\includegraphics[scale=0.25,trim = 34cm 1cm 0 0, clip=true]{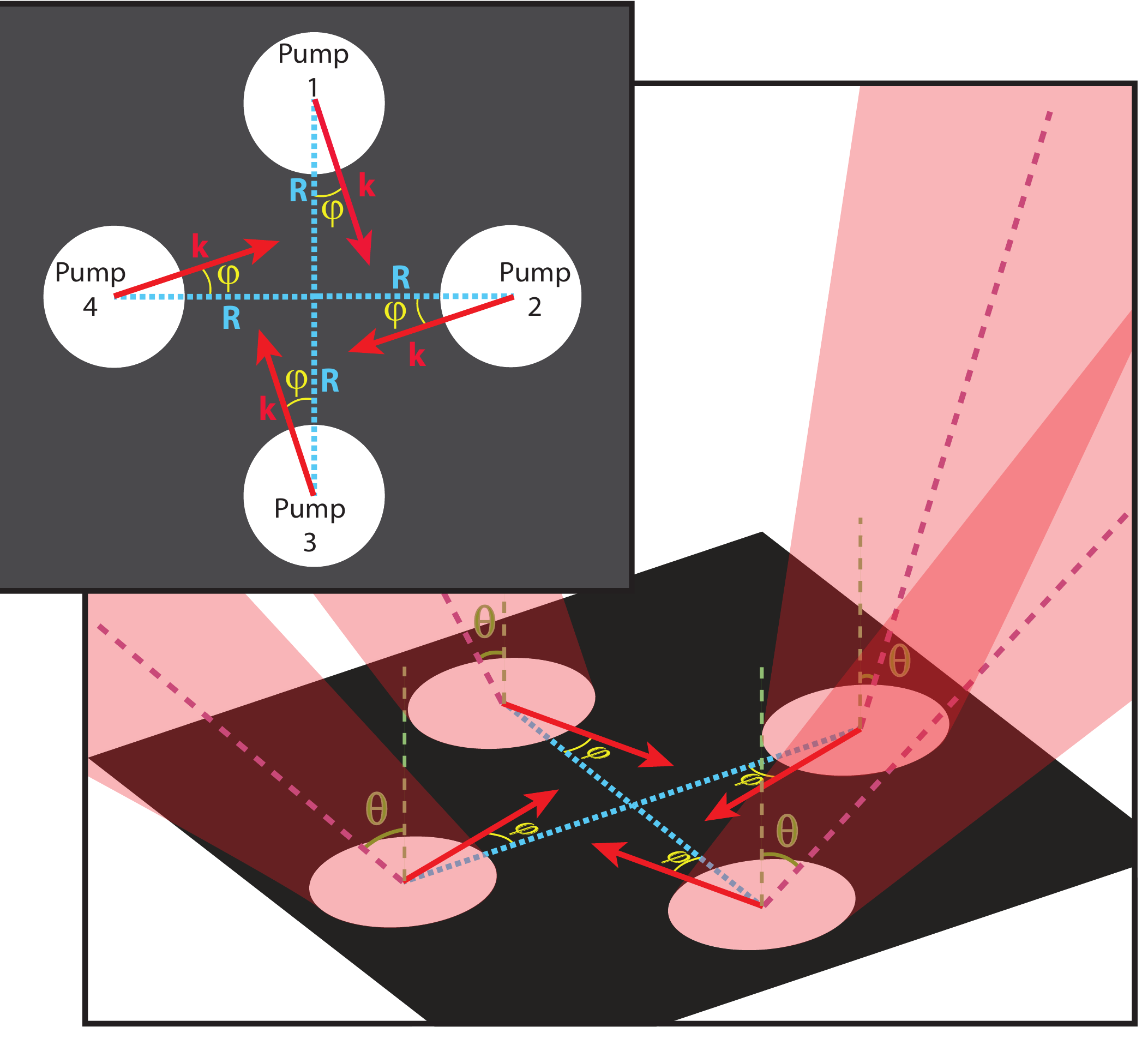}
\caption{Schematic representation of the four pumps arriving on the sample.
The laser beam is focused on a pinhole before being divided into four {equal beams}. They are focused on the sample {($a=\unit{12}{\micro\meter}$ waist)} so that they form four polariton fluids propagating towards each other.
$\theta$ is the incidence angle, giving the norm of the polariton wavevector~$\mathbf{k}$. The in-plane polariton propagation direction is set {by} the azimuthal angle~$\varphi$. The blue dashed lines show the direction to the center ($\varphi=0$) while the red arrows show the polariton direction of propagation {for~$\theta\not= 0$}. $R$~is the distance of the pumps to the center ($R=\unit{25(3)}{\micro\metre}$).}
\label{Setup4pumps}
\end{center}
\end{figure}

In the present work, we use four laser beams arranged in a square to resonantly inject polaritons going towards the center of the square. By slightly tilting the pumping direction of the laser beams the four convergent polariton populations can be made to propagate with a small angle {with the direction to the center} (see~Fig.\,\ref{Setup4pumps}), {therefore injecting} a controlled angular momentum into the polariton fluid. 
At the same time the resonant pumping configuration allows a fine tuning of the polariton density without generating an excitonic reservoir, and consequently a precise control of the nonlinearities in the system, in contrast with the case of out-of-resonant schemes~\cite{Dall2014, tosi12, cristofolini13}.

With this setup, we demonstrate a new technique for the injection of topological charges in polaritons superfluids, a problem of major relevance in driven-dissipative open systems strongly coupled to the environment.
Moreover, our study indicates that, in the steady-state regime, the angular momentum continuously injected by the pumps compensates the loss of angular momentum by the system.

\paragraph{Experiment.}
\addcontentsline{toc}{section}{Experiment}

 The sample is a $2\lambda$-GaAs planar microcavity containing three ${\text{In}}_{0.05}{\text{Ga}}_{0.95}\text{As}$ quantum wells. {The finesse} is about~$3000$, which amounts to a polariton linewidth smaller than $\unit{0.1}{\milli\electronvolt}$ and a Rabi spliting~$\Omega_\text{R}=\unit{5.1}{\milli\electronvolt}$. 
 A wedge between the two cavity Bragg mirrors allows controlling the photon-exciton energy detuning at normal incidence~$\delta$, by choosing the appropriate region on the cavity.
 All measurements presented here were done in a region where the natural cavity disorder is minimum.
  This region has a slightly positive detuning ($\delta=+\unit{0.5}{\milli\electronvolt}$), which provides a good balance between strong interactions and long propagation distance. 
 The lower-polariton resonance at $\abs{\mathbf{k}}=0$ and $\delta=0$ is at $\unit{837}{\nano\meter}$. 
 The cavity is pumped resonantly with a single mode CW Ti:Sa laser, frequency-locked to an optical cavity.

The laser is spatially filtered by a~$\unit{50}{\micro\meter}$ pinhole so that the Gaussian tail is cut to minimize pumps overlap. Three beamsplitters divide the laser into four beams of equal intensities, as shown in figure~\ref{Setup4pumps}. The four arms are sent along similar trajectories by dielectric mirrors.
Each laser beam is then focused on the sample with a single condenser lens. The pumps are circularly polarized by a quarter-wave plate before hitting the sample. In this way we obtain only one kind of polariton population and avoid any effect due to spin-dependent interactions~\cite{vladimirova10}. For each pump, the real space positions and the angle of incidence can be controlled independently.

The four resonant pumps are spatially arranged on the sample to form a square and are described by $F(\mathbf{r})=\sum_{j=1}^4 F_j(\mathbf{r})\exp{(-i\,\mathbf{k}_j \cdot \mathbf{r})}$, where the $F_j(\mathbf{r})$ are the four spatial profiles. 
Their position in k-space is chosen so that polaritons from each pump propagate towards the square center. The four in-plane wavevectors are chosen with the same norm~$|\mathbf{k}_j|=|\mathbf{k}|$, meaning that all four pumps hit the sample with the same angle of incidence~$\theta$. 
For a fixed~$\theta$, we tilt the in-plane direction of propagation by an angle $\varphi$ relative to the direction of the center (see~Fig.~\ref{Setup4pumps}). 
This allows sending onto the cavity a continuous orbital angular momentum (OAM) per photon~\cite{Amaral:14,Allen200067}, in unit of~$\hbar$ that can be evaluated as:
\begin{equation}
	\frac{L}{\hbar} = 
	\frac{1}{\mathcal{N}}
	\iint \mathrm{d}{x}\,\mathrm{d}{y} \, 
	F^*(\mathbf{r})
	\hat{L}_z
	F(\mathbf{r}) = R |\mathbf{k}| \sin\varphi,
	\label{ell_4pumps}
\end{equation}
where $\hat{L}_z=\hbar\left(x\frac{\partial}{\partial y}-y\frac{\partial}{\partial x}\right)$ is the~$z$ component of the angular momentum, $R$ is the pump distance to the square center and $\mathcal{N}=\iint \, \mathrm{d}\mathbf{r} \, F^*(\mathbf{r})F(\mathbf{r})$~is the normalization constant. Equation~(\ref{ell_4pumps}) has been derived considering a perfectly symmetric system and non-overlapping pumps with circular profiles induced by the pinhole.
As shown in the~Supplementary material, in the steady-state regime, the average angular momentum per photon injected by the pumps is equal to the average angular momentum per polariton inside the cavity.
Note that since the spatial independance of the injected fields lifts the constrain of the phase circulation quantization, a non-integer (\textit{i.e.{}}~real-valued) OAM can be injected, which is impossible for a single~Laguerre-Gauss field. 

The separation between the pumps is small enough so that the four polariton populations can meet, resulting in a significant density at the square center. 
Cutting the beam Gaussian tails results in negligible direct illumination in the central region of the square. This ensures that in the central region polaritons are free to evolve. If both~$\theta$ and~$\varphi$ are non-zero, the four polariton populations meet in the system center and an angular momentum is injected.

\begin{figure*}[t!]
	\includegraphics[width=0.85\textwidth]{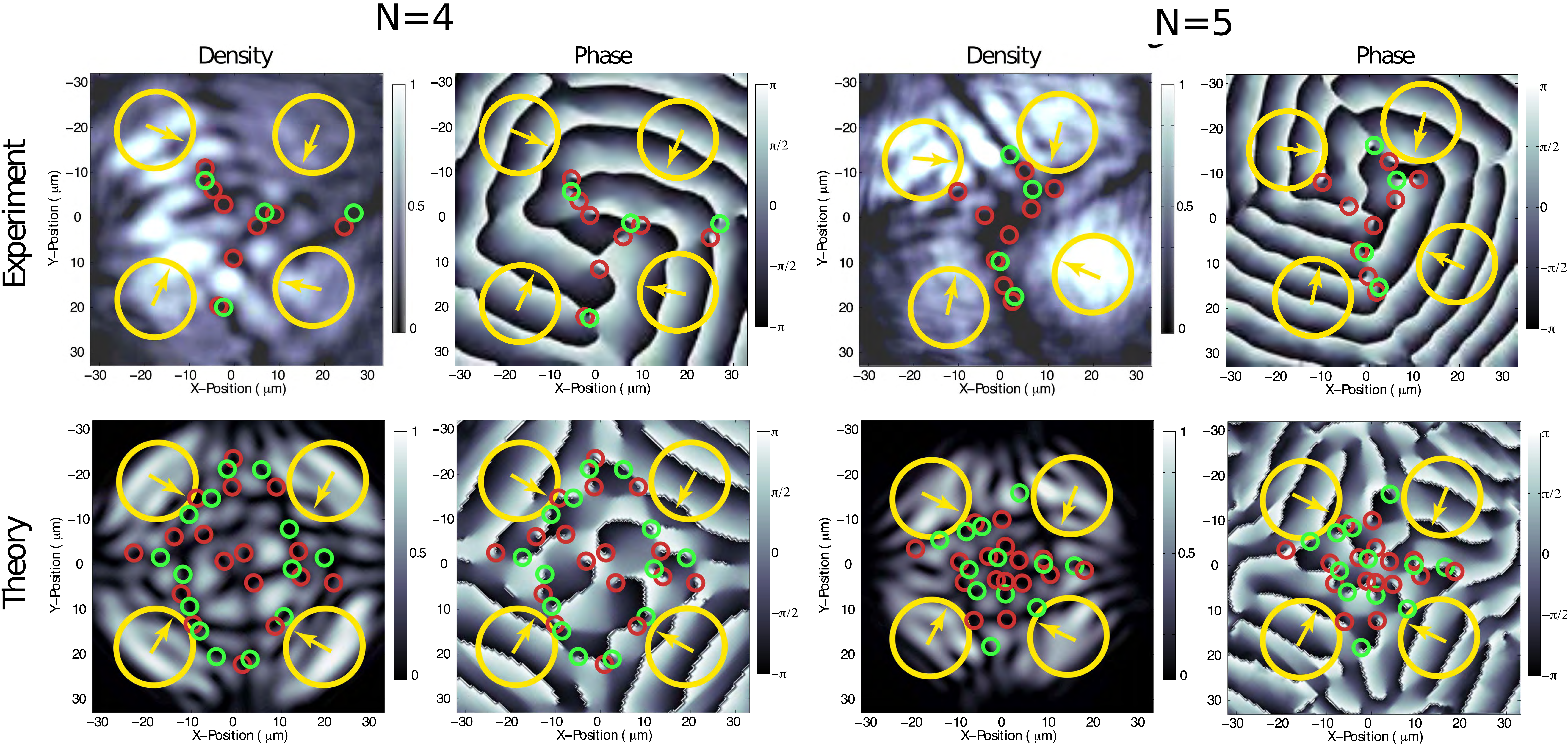}
	\caption{Experimental (up) and theoretical (down) density and phase maps for $\theta=3.5\degree$, $\varphi=\unit{21}{\degree}$ ({$L=4.0(5)$}) and $\varphi=\unit{26}{\degree}$ ({$L=4.9(6)$}) in the low density regime. An interference pattern is visible together with phase singularities of both signs. Since the injected angular momentum is not zero the number of ($+$) singularities (red circles) is different from the number of ($-$) singularities (green circles).}
	\label{4PumpslowDTilts}
\end{figure*}

An objective collects the sample emission and the time-averaged detection is made simultaneously through direct imaging with~CCD cameras in real space and momentum space. The energy is measured with a spectrometer. We only collect circularly-polarized light, therefore filtering out any spin-flip effect.
 The polariton phase is measured with an off-axis interferometry setup: a beam splitter divides the real space image into two parts, one of which is expanded to generate a flat phase reference beam, used to make an off-axis interference pattern. With this method, the vortex position on the image is independent of the phase of the reference beam~\cite{Bolda98}.
 The actual phase map is then numerically reconstructed with a phase retrieval algorithm.

\paragraph{Numerical method.}
\addcontentsline{toc}{section}{Numerical method}
To describe the configuration under study, we numerically solve the driven-dissipative scalar Gross-Pitaevskii equation.
 The field variable $\psi \equiv \psi(\mathbf{r}) = \big\langle\hat{\psi}(\mathbf{r})\big\rangle$ is the mean value of the real-space polariton field operator $\hat{\psi}(\mathbf{r})$.
This equation describes the mean field for bi-dimensional interacting particles with a pump and a decay term as
\begin{equation}
i\hbar \frac{\partial \psi}{\partial t}=\left(-\frac{\hbar^2\nabla^2}{2m^*}-\frac{i\hbar \gamma}{2}+ g \vert \psi\vert^2 \right)\psi
+\hbar\gamma\,F(\mathbf{r})e^{i\Delta t},
\label{GP}
\end{equation}
where $m^*$ is the polariton effective mass equal to~$9.7\cdot10^{-5}$ the electron mass, $\gamma$ is the decay rate deduced from the polariton lifetime (here $1/\gamma=\unit{12}{p\second}$), $g=\unit{5}{\micro\electronvolt \micro\meter^2}$ is the polariton-polariton interaction and~$\Delta=\omega_\mathrm{l}-\omega_\mathrm{LP}(|\mathbf{k}|)$ is the energy detuning between the pump laser frequency ($\omega_\mathrm{l}$) and the lower polariton branch at~$|\mathbf{k}|$, here $\unit{0.3}{\milli\electronvolt}$. This detuning allows to compensate for the shift of the lower polariton branch appearing at higher intensities and to have high densities in the superfluid regime.
Direct comparison with the experiment is  performed by extracting the steady-state density $\abs{\psi}^2$ and phase $\arg(\psi)$. 
In these conditions, the simulations give a non-turbulent steady-state regime. 
Therefore, the resulting density and phase maps shown in figure~\ref{4PumpslowDTilts} are equivalent to time-averaged maps, as in the experiment.

\paragraph{Results.}
\addcontentsline{toc}{section}{Results}
To highlight the role of polariton--polariton interactions, we study the system as a function of the polariton density. We identify two different regimes: a linear regime at low polariton density, where interactions can be neglected, and a nonlinear regime at high  density, were polaritons have a superfluid behavior~\cite{Amo09a, Amo11, Hivet12}.
Moreover, in the superfluid state we observe the vanishing of the interference visible in the linear regime when two or more fluids meet~\cite{Boulier15}. This phenomenon was predicted and observed to be accompanied by the annihilation of all vortex-antivortex (V-AV) pairs~\cite{Cancellieri14, hivet2014interaction}.

\begin{figure*}
\begin{center}
\includegraphics[width=0.85\textwidth]{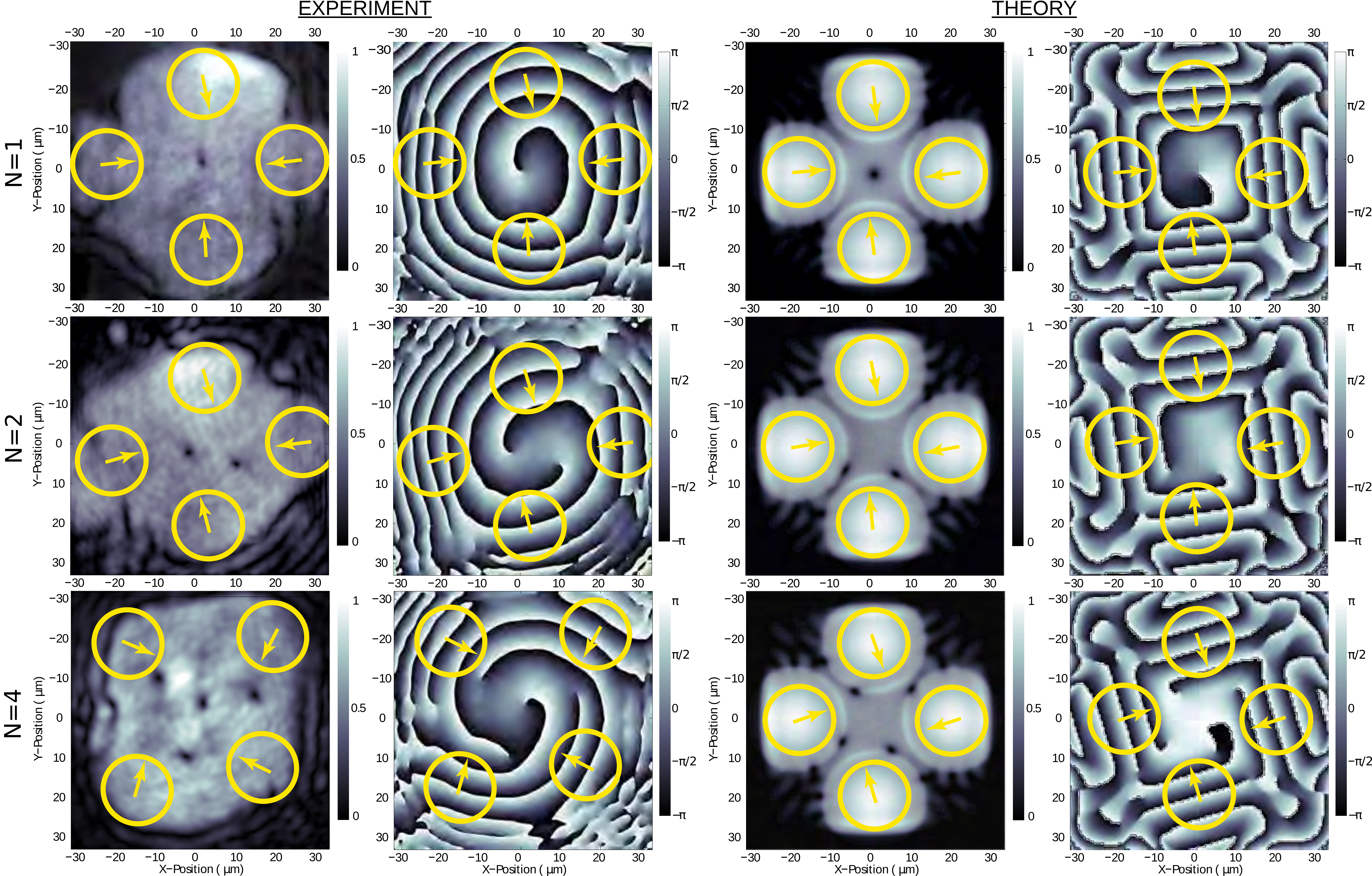}
\caption{Experimental (left) and theoretical (right) density and phase maps for~{$L=1.1(1), 2.0(2), 4.0(5)$} (from top to bottom) at high density. The vortex number~$N$ is equal to the integer part of~$L$. The vortices are visible as black dots in the density, each associated with a phase singularity. On average~$\abs{\mathbf{k}}=\unit{0.45}{\micro\meter^{-1}}$ and (from top to bottom) $\varphi=\unit{5.5}{\degree},\unit{10}{\degree},\unit{21}{\degree}$.}
\label{4PumpsHigDTilt}
\end{center}
\end{figure*}

For low densities, as shown in figure~\ref{4PumpslowDTilts}, a square interference pattern is visible. This is the behavior expected for non-interacting polaritons, which are similar to cavity photons.
Moreover phase singularities of both signs are visible. It is important to note that in the linear regime, no healing length can be defined in the density. 
Therefore the hydrodynamic definition of a vortex core cannot be applied. 
In this regimes, we observe an unequal number of singularities of opposite signs.
The difference between the number of vortices and anti-vortices ($N\equiv N_+-N_-$) is equal to the integer part of the angular momentum~{$L$} expected from equation~\eqref{ell_4pumps}. This shows that our technique allows the injection of topological charges by means of OAM. These observations are in agreement with the fact that the sample disorder only generates V-AV pairs~\cite{Liew08,Nardin11,Lagoudakis08,Pigeon11}.
Figure~\ref{4PumpslowDTilts} gives an example of low density regime for $\varphi=\unit{21}{\degree}$ and $\varphi=\unit{26}{\degree}$ for a fixed incident angle~$\theta=\unit{3.5}{\degree}$ ($\abs{\mathbf{k}}=\unit{0.45}{\micro\meter^{-1}}$), corresponding to $L=4.0(5)$ and $L=4.9(6)$ respectively (uncertainty based on the error in the evaluation of~$R$). Equation~\eqref{GP} provides qualitatively correct predictions in the linear regime. 
A difference in the number of V-AV pairs between experiment and simulation can be ascribed to imperfections of the sample, which at low density play an important role. In this regime the system reaches a steady state with vortices lying in fixed positions. This is confirmed by the simulations, as said before, and by high values of visibility (not shown here) all over the pumped region, apart from the vortex cores.

\begin{figure}
	\begin{center}
	\includegraphics[width=0.5\textwidth]{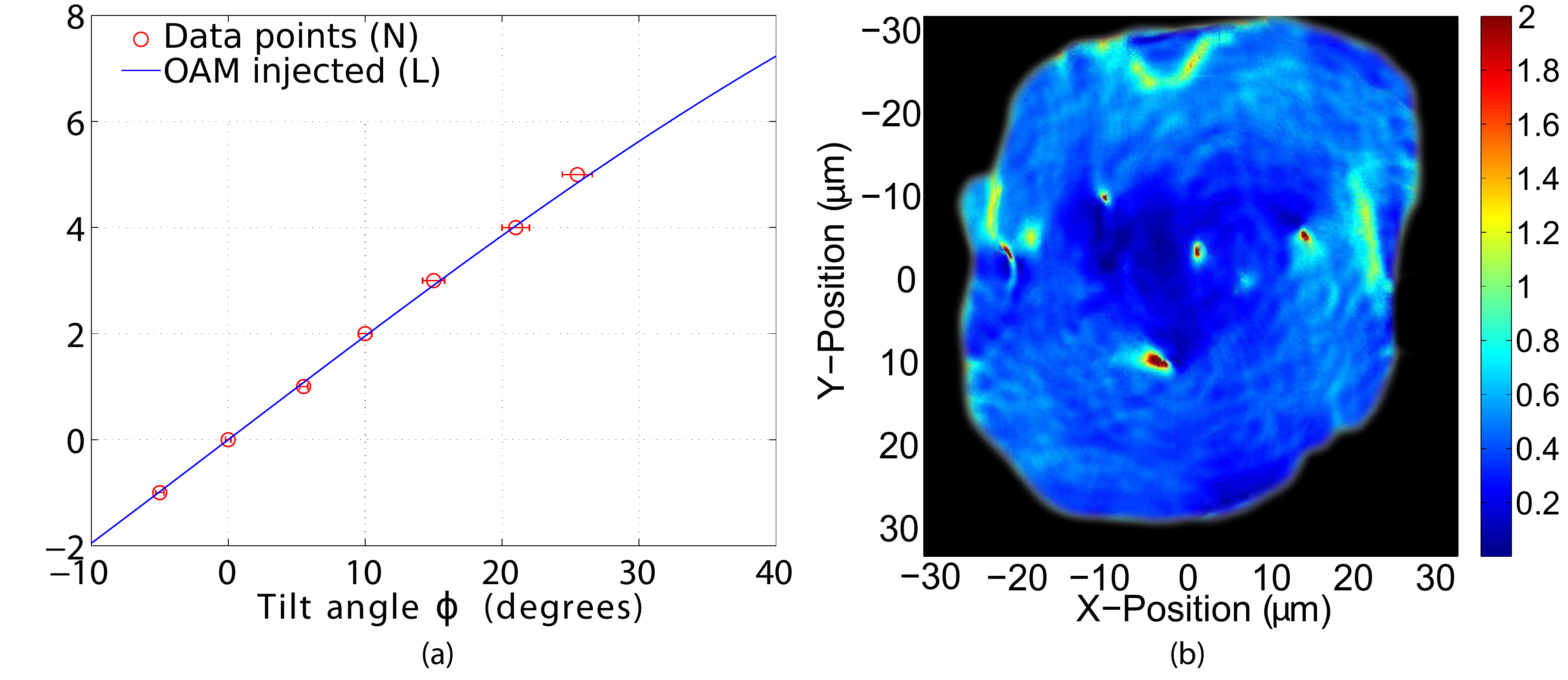}
	\caption{(a)~Plot of the observed number of vortices~$N$ (red circles) and of the continuous angular momentum~$L$ from equation~\eqref{ell_4pumps} (blue line) as a fonction of the azimuthal angle~$\varphi$.
(b)~Experimental map of the Mach number for the case of~$L=4.0(5)$. The black zones signify areas outside of the superfluid, where the polariton density is too low to define a Mach number. In the blue zone, the fluid is subsonic. However, polaritons within each vortex are strongly supersonic. Note that the Mach number scale is limited between~0 and~2 while at the vortex core the Mach number diverges and experimental values up to 100 are obtained.}
	\label{EllVsTilt}
	\end{center}
\end{figure}

By increasing the density to the point where, in the central region, the polariton fluid reaches the superfluid regime (see Fig.\,\ref{4PumpsHigDTilt}), the interference pattern disappears and all V-AV pairs annihilate, showing the interaction-driven nature of this phenomenon~\cite{Cancellieri14, Boulier15}. 
The nonzero angular momentum injected by the pumps results in the presence of elementary vortices of the same sign remaining in the superfluid.
Their size is of the order of the healing length (about~$\unit{2}{\micro\metre}$) that can be unambiguously defined~\cite{Boulier15, Amo09a, Amo09b}.
Up to five vortices were observed without any antivortex. Figure~\ref{4PumpsHigDTilt} shows the experimental results for~$\varphi=\unit{5.5}{\degree}$ ({$L=1.1(1)$}), $\varphi=\unit{10}{\degree}$ ({$L=2.0(2)$}) and $\varphi=\unit{21}{\degree}$ ({$L=4.0(5)$}), for a fixed $\theta=\unit{3.5}{\degree}$ giving $\abs{\mathbf{k}}=\unit{0.45}{\micro\meter^{-1}}$.
As expected, the number of vortices increases with $\varphi$.

A comparison between the observed number of vortices~$N$ in the superfluid regime and the value of~{$L$} computed from the classical approach~\eqref{ell_4pumps} is presented in figure~\ref{EllVsTilt}a. The agreement is good within the uncertainty on~$\varphi$ and~$R$, showing the validity of our approach that allows the storage of quantized vortices by injecting orbital angular momentum.
The vortex position is observed to depend strongly on each pump phase, which suggests that the vortex lattice shape is related to the geometry of the polariton superflow. This, together with the presence of disorder in the sample, can explain the discrepancies in the vortex position in the model and in the experiment. In the numerical simulations, the pumps phase and position are set slightly different in order to reproduce the asymmetry of the experimental case.

Finally, in the high density case, it is interesting to look at the~Mach number map which is defined as the ratio between the local velocity of the fluid and the local speed of sound (proportional to the square root of the polariton density). Figure~\ref{EllVsTilt}b shows the Mach number map for $L=4.0(5)$ corresponding to the bottom panels of figure~\ref{4PumpsHigDTilt}. As expected from the absence of interference pattern, the fluid is subsonic ($M<1$), which means that it is in the superfluid regime \cite{Amo09b,Amo09a}. Note however that polaritons within each vortex are strongly supersonic. Indeed, at the vortex core the Mach number is expected to diverge, and experimental values up to~$M=100$ are obtained.

\paragraph{Conclusion.}
\addcontentsline{toc}{section}{Conclusion}
 We designed a scheme that allows the injection of a controlled angular momentum in a bi-dimensional polariton superfluid. The scheme makes use of four coherent polariton populations. While in the linear regime interferences appear and phase singularities of opposite charges are clearly visible, the vanishing of all possible pairs happens at sufficiently high density, and only same sign vortices survive as expected for a coherent superfluid.
 We therefore observed the injection of angular momentum and the storage of topological charges in a non-equilibrium superfluid of light. 
 Together with the recent result obtained in {reference}~\cite{Boulier15}, this is an interesting achievement in polariton physics. Up to now only vortex-antivortex pairs~\cite{Nardin11, Lagoudakis08, tosi12, roumpos11, manni13}, single vortex \cite{dominici14, sanvitto10} and vortices with high~{$L$} confined by an excitonic reservoir~\cite{Dall2014} have been observed.
Our method allows to very efficiently imprint large values of orbital angular momentum and observe several vortices with a topological charge of~$1$. With a large number of vortices, this result opens the way to the study of vortex-vortex interactions, vortex lattices and their collective modes~\cite{tkachenko66,coddington2003}. The technique presented in this paper, coupled with  the use of a time-resolved set-up, could also lead to interesting new studies in the physics of vortices in a turbulent regime~\cite{Nardin11}.

\textbf{{Acknowledgments}}
We acknowledge the financial support of the ANR Quandyde (ANR-11-BS10-001) and EPSRC grant EP/J007544, and Iacopo Carusotto and Paulo Maia Neto for very helpfull discussions.


\begin{thebibliography}{10}
\expandafter\ifx\csname url\endcsname\relax
  \def\url#1{\texttt{#1}}\fi
\expandafter\ifx\csname urlprefix\endcsname\relax\def\urlprefix{URL }\fi
\providecommand{\bibinfo}[2]{#2}
\providecommand{\eprint}[2][]{\url{#2}}

\bibitem{Weisbuch92}
\bibinfo{author}{Weisbuch, C.}, \bibinfo{author}{Nishioka, M.},
  \bibinfo{author}{Ishikawa, A.} \& \bibinfo{author}{Arakawa, Y.}
\newblock \bibinfo{title}{Observation of the coupled exciton-photon mode
  splitting in a semiconductor quantum microcavity}.
\newblock \emph{\bibinfo{journal}{Phys. Rev. Lett.}}
  \textbf{\bibinfo{volume}{69}}, \bibinfo{pages}{3314} (\bibinfo{year}{1992}).

\bibitem{RevModPhys.82.1489}
\bibinfo{author}{Deng, H.}, \bibinfo{author}{Haug, H.} \&
  \bibinfo{author}{Yamamoto, Y.}
\newblock \bibinfo{title}{Exciton-polariton {B}ose-{E}instein condensation}.
\newblock \emph{\bibinfo{journal}{Rev. Mod. Phys.}}
  \textbf{\bibinfo{volume}{82}}, \bibinfo{pages}{1489--1537}
  (\bibinfo{year}{2010}).

\bibitem{RevModPhys.85.299}
\bibinfo{author}{Carusotto, I.} \& \bibinfo{author}{Ciuti, C.}
\newblock \bibinfo{title}{Quantum fluids of light}.
\newblock \emph{\bibinfo{journal}{Rev. Mod. Phys.}}
  \textbf{\bibinfo{volume}{85}}, \bibinfo{pages}{299--366}
  (\bibinfo{year}{2013}).

\bibitem{RevModPhys.71.463}
\bibinfo{author}{Dalfovo, F.}, \bibinfo{author}{Giorgini, S.},
  \bibinfo{author}{Pitaevskii, L.~P.} \& \bibinfo{author}{Stringari, S.}
\newblock \bibinfo{title}{Theory of {B}ose-{E}instein condensation in trapped
  gases}.
\newblock \emph{\bibinfo{journal}{Rev. Mod. Phys.}}
  \textbf{\bibinfo{volume}{71}}, \bibinfo{pages}{463--512}
  (\bibinfo{year}{1999}).

\bibitem{PhysRevLett.86.4447}
\bibinfo{author}{Burger, S.} \emph{et~al.}
\newblock \bibinfo{title}{Superfluid and dissipative dynamics of a
  {B}ose-{E}instein condensate in a periodic optical potential}.
\newblock \emph{\bibinfo{journal}{Phys. Rev. Lett.}}
  \textbf{\bibinfo{volume}{86}}, \bibinfo{pages}{4447--4450}
  (\bibinfo{year}{2001}).

\bibitem{PhysRevLett.93.166401}
\bibinfo{author}{Carusotto, I.} \& \bibinfo{author}{Ciuti, C.}
\newblock \bibinfo{title}{Probing microcavity polariton superfluidity through
  resonant rayleigh scattering}.
\newblock \emph{\bibinfo{journal}{Phys. Rev. Lett.}}
  \textbf{\bibinfo{volume}{93}}, \bibinfo{pages}{166401}
  (\bibinfo{year}{2004}).

\bibitem{Amo09b}
\bibinfo{author}{Amo, A.} \emph{et~al.}
\newblock \bibinfo{title}{Collective fluid dynamics of a polariton condensate
  in a semiconductor microcavity}.
\newblock \emph{\bibinfo{journal}{Nature}} \textbf{\bibinfo{volume}{457}},
  \bibinfo{pages}{291--295} (\bibinfo{year}{2009}).

\bibitem{Amo09a}
\bibinfo{author}{Amo, A.} \emph{et~al.}
\newblock \bibinfo{title}{Superfluidity of polaritons in semiconductor
  microcavities}.
\newblock \emph{\bibinfo{journal}{Nature Phys.}} \textbf{\bibinfo{volume}{5}},
  \bibinfo{pages}{805--810} (\bibinfo{year}{2009}).

\bibitem{sanvitto10}
\bibinfo{author}{Sanvitto, D.} \emph{et~al.}
\newblock \bibinfo{title}{Persistent currents and quantized vortices in a
  polariton superfluid}.
\newblock \emph{\bibinfo{journal}{Nature Phys.}} \textbf{\bibinfo{volume}{6}},
  \bibinfo{pages}{527--533} (\bibinfo{year}{2010}).

\bibitem{desyatnikov2005optical}
\bibinfo{author}{Desyatnikov, A.~S.}, \bibinfo{author}{Torner, L.} \&
  \bibinfo{author}{Kivshar, Y.~S.}
\newblock \bibinfo{title}{Optical vortices and vortex solitons}.
\newblock \emph{\bibinfo{journal}{Progress in Optics}}
  \textbf{\bibinfo{volume}{47}}, \bibinfo{pages}{291--391}
  (\bibinfo{year}{2005}).

\bibitem{essmann1967direct}
\bibinfo{author}{Essmann, U.} \& \bibinfo{author}{Tr{\"a}uble, H.}
\newblock \bibinfo{title}{The direct observation of individual flux lines in
  type ii superconductors}.
\newblock \emph{\bibinfo{journal}{Phys. Lett. A}}
  \textbf{\bibinfo{volume}{24}}, \bibinfo{pages}{526--527}
  (\bibinfo{year}{1967}).

\bibitem{PhysRev.136.A1194}
\bibinfo{author}{Rayfield, G.~W.} \& \bibinfo{author}{Reif, F.}
\newblock \bibinfo{title}{Quantized vortex rings in superfluid helium}.
\newblock \emph{\bibinfo{journal}{Phys. Rev.}} \textbf{\bibinfo{volume}{136}},
  \bibinfo{pages}{A1194--A1208} (\bibinfo{year}{1964}).

\bibitem{madison2000vortex}
\bibinfo{author}{Madison, K.~W.}, \bibinfo{author}{Chevy, F.},
  \bibinfo{author}{Wohlleben, W.} \& \bibinfo{author}{Dalibard, J.}
\newblock \bibinfo{title}{Vortex formation in a stirred {B}ose-{E}instein
  condensate}.
\newblock \emph{\bibinfo{journal}{Phys. Rev. Lett.}}
  \textbf{\bibinfo{volume}{84}}, \bibinfo{pages}{806} (\bibinfo{year}{2000}).

\bibitem{denschlag2000generating}
\bibinfo{author}{Denschlag, J.} \emph{et~al.}
\newblock \bibinfo{title}{Generating solitons by phase engineering of a
  {B}ose-{E}instein condensate}.
\newblock \emph{\bibinfo{journal}{Science}} \textbf{\bibinfo{volume}{287}},
  \bibinfo{pages}{97--101} (\bibinfo{year}{2000}).

\bibitem{khaykovich2002formation}
\bibinfo{author}{Khaykovich, L.} \emph{et~al.}
\newblock \bibinfo{title}{Formation of a matter-wave bright soliton}.
\newblock \emph{\bibinfo{journal}{Science}} \textbf{\bibinfo{volume}{296}},
  \bibinfo{pages}{1290--1293} (\bibinfo{year}{2002}).

\bibitem{Pigeon11}
\bibinfo{author}{Pigeon, S.}, \bibinfo{author}{Carusotto, I.} \&
  \bibinfo{author}{Ciuti, C.}
\newblock \bibinfo{title}{Hydrodynamic nucleation of vortices and solitons in a
  resonantly excited polariton superfluid}.
\newblock \emph{\bibinfo{journal}{Phys. Rev. B}} \textbf{\bibinfo{volume}{83}},
  \bibinfo{pages}{144513} (\bibinfo{year}{2011}).

\bibitem{roumpos11}
\bibinfo{author}{Roumpos, G.} \emph{et~al.}
\newblock \bibinfo{title}{Single vortex-antivortex pair in an exciton-polariton
  condensate}.
\newblock \emph{\bibinfo{journal}{Nature Phys.}} \textbf{\bibinfo{volume}{7}},
  \bibinfo{pages}{129--133} (\bibinfo{year}{2011}).

\bibitem{Nardin11}
\bibinfo{author}{Nardin, G.} \emph{et~al.}
\newblock \bibinfo{title}{Hydrodynamic nucleation of quantized vortex pairs in
  a polariton quantum fluid}.
\newblock \emph{\bibinfo{journal}{Nature Phys.}} \textbf{\bibinfo{volume}{7}},
  \bibinfo{pages}{635--641} (\bibinfo{year}{2011}).

\bibitem{sanvitto11}
\bibinfo{author}{Sanvitto, D.} \emph{et~al.}
\newblock \bibinfo{title}{All-optical control of the quantum flow of a
  polariton condensate}.
\newblock \emph{\bibinfo{journal}{Nature Photon.}}
  \textbf{\bibinfo{volume}{5}}, \bibinfo{pages}{610--614}
  (\bibinfo{year}{2011}).

\bibitem{Lagoudakis08}
\bibinfo{author}{Lagoudakis, K.~G.} \emph{et~al.}
\newblock \bibinfo{title}{Quantized vortices in an exciton--polariton
  condensate}.
\newblock \emph{\bibinfo{journal}{Nature Phys.}} \textbf{\bibinfo{volume}{4}},
  \bibinfo{pages}{706--710} (\bibinfo{year}{2008}).

\bibitem{manni13}
\bibinfo{author}{Manni, F.} \emph{et~al.}
\newblock \bibinfo{title}{Spontaneous self-ordered states of vortex-antivortex
  pairs in a polariton condensate}.
\newblock \emph{\bibinfo{journal}{Phys. Rev. B}} \textbf{\bibinfo{volume}{88}},
  \bibinfo{pages}{201303} (\bibinfo{year}{2013}).

\bibitem{zhenghan10}
\bibinfo{author}{Zhenghan, W.}
\newblock \emph{\bibinfo{title}{Topological quantum computation}}.
\newblock \bibinfo{number}{112} (\bibinfo{publisher}{American Mathematical
  Soc.}, \bibinfo{year}{2010}).

\bibitem{freedman03}
\bibinfo{author}{Freedman, M.}, \bibinfo{author}{Kitaev, A.},
  \bibinfo{author}{Larsen, M.} \& \bibinfo{author}{Wang, Z.}
\newblock \bibinfo{title}{Topological quantum computation}.
\newblock \emph{\bibinfo{journal}{Bulletin of the American Mathematical
  Society}} \textbf{\bibinfo{volume}{40}}, \bibinfo{pages}{31--38}
  (\bibinfo{year}{2003}).

\bibitem{nayak08}
\bibinfo{author}{Nayak, C.}, \bibinfo{author}{Simon, S.~H.},
  \bibinfo{author}{Stern, A.}, \bibinfo{author}{Freedman, M.} \&
  \bibinfo{author}{Sarma, S.~D.}
\newblock \bibinfo{title}{Non-abelian anyons and topological quantum
  computation}.
\newblock \emph{\bibinfo{journal}{Reviews of Modern Physics}}
  \textbf{\bibinfo{volume}{80}}, \bibinfo{pages}{1083} (\bibinfo{year}{2008}).

\bibitem{Dall2014}
\bibinfo{author}{Dall, R.} \emph{et~al.}
\newblock \bibinfo{title}{Creation of orbital angular momentum states with
  chiral polaritonic lenses}.
\newblock \emph{\bibinfo{journal}{Phys. Rev. Lett.}}
  \textbf{\bibinfo{volume}{113}}, \bibinfo{pages}{200404}
  (\bibinfo{year}{2014}).

\bibitem{lagoudakis09}
\bibinfo{author}{Lagoudakis, K.} \emph{et~al.}
\newblock \bibinfo{title}{Observation of half-quantum vortices in an
  exciton-polariton condensate}.
\newblock \emph{\bibinfo{journal}{Science}} \textbf{\bibinfo{volume}{326}},
  \bibinfo{pages}{974--976} (\bibinfo{year}{2009}).

\bibitem{flayac10}
\bibinfo{author}{Flayac, H.}, \bibinfo{author}{Shelykh, I.~A.},
  \bibinfo{author}{Solnyshkov, D.~D.} \& \bibinfo{author}{Malpuech, G.}
\newblock \bibinfo{title}{Topological stability of the half-vortices in spinor
  exciton-polariton condensates}.
\newblock \emph{\bibinfo{journal}{Phys. Rev. B}} \textbf{\bibinfo{volume}{81}},
  \bibinfo{pages}{045318} (\bibinfo{year}{2010}).

\bibitem{tosi11}
\bibinfo{author}{Tosi, G.} \emph{et~al.}
\newblock \bibinfo{title}{Onset and dynamics of vortex-antivortex pairs in
  polariton optical parametric oscillator superfluids}.
\newblock \emph{\bibinfo{journal}{Phys. Rev. Lett.}}
  \textbf{\bibinfo{volume}{107}}, \bibinfo{pages}{036401}
  (\bibinfo{year}{2011}).

\bibitem{spin-vortices15}
\bibinfo{author}{Dufferwiel, S.} \emph{et~al.}
\newblock \bibinfo{title}{Spin textures of exciton-polaritons in a tunable
  microcavity with large te-tm splitting}.
\newblock \emph{\bibinfo{journal}{Phys. Rev. Lett.}}
  \textbf{\bibinfo{volume}{115}}, \bibinfo{pages}{246401}
  (\bibinfo{year}{2015}).

\bibitem{keeling08}
\bibinfo{author}{Keeling, J.} \& \bibinfo{author}{Berloff, N.~G.}
\newblock \bibinfo{title}{Spontaneous rotating vortex lattices in a pumped
  decaying condensate}.
\newblock \emph{\bibinfo{journal}{Phys. Rev. Lett.}}
  \textbf{\bibinfo{volume}{100}}, \bibinfo{pages}{250401}
  (\bibinfo{year}{2008}).

\bibitem{Gorbach10}
\bibinfo{author}{Gorbach, A.~V.}, \bibinfo{author}{Hartley, R.} \&
  \bibinfo{author}{Skryabin, D.~V.}
\newblock \bibinfo{title}{Vortex lattices in coherently pumped polariton
  microcavities}.
\newblock \emph{\bibinfo{journal}{Phys. Rev. Lett.}}
  \textbf{\bibinfo{volume}{104}}, \bibinfo{pages}{213903}
  (\bibinfo{year}{2010}).

\bibitem{Liew08}
\bibinfo{author}{Liew, T. C.~H.}, \bibinfo{author}{Rubo, Y.~G.} \&
  \bibinfo{author}{Kavokin, A.~V.}
\newblock \bibinfo{title}{Generation and dynamics of vortex lattices in
  coherent exciton-polariton fields}.
\newblock \emph{\bibinfo{journal}{Phys. Rev. Lett.}}
  \textbf{\bibinfo{volume}{101}}, \bibinfo{pages}{187401}
  (\bibinfo{year}{2008}).

\bibitem{Kusudo13}
\bibinfo{author}{Kusudo, K.} \emph{et~al.}
\newblock \bibinfo{title}{Stochastic formation of polariton condensates in two
  degenerate orbital states}.
\newblock \emph{\bibinfo{journal}{Phys. Rev. B}} \textbf{\bibinfo{volume}{87}},
  \bibinfo{pages}{214503} (\bibinfo{year}{2013}).

\bibitem{tosi12}
\bibinfo{author}{Tosi, G.} \emph{et~al.}
\newblock \bibinfo{title}{Geometrically locked vortex lattices in semiconductor
  quantum fluids}.
\newblock \emph{\bibinfo{journal}{Nature Commun..}}
  \textbf{\bibinfo{volume}{3}}, \bibinfo{pages}{1243} (\bibinfo{year}{2012}).

\bibitem{cristofolini13}
\bibinfo{author}{Cristofolini, P.} \emph{et~al.}
\newblock \bibinfo{title}{Optical superfluid phase transitions and trapping of
  polariton condensates}.
\newblock \emph{\bibinfo{journal}{Phys. Rev. Lett.}}
  \textbf{\bibinfo{volume}{110}}, \bibinfo{pages}{186403}
  (\bibinfo{year}{2013}).

\bibitem{vladimirova10}
\bibinfo{author}{Vladimirova, M.} \emph{et~al.}
\newblock \bibinfo{title}{Polariton-polariton interaction constants in
  microcavities}.
\newblock \emph{\bibinfo{journal}{Phys. Rev. B}} \textbf{\bibinfo{volume}{82}},
  \bibinfo{pages}{075301} (\bibinfo{year}{2010}).

\bibitem{Amaral:14}
\bibinfo{author}{Amaral, A.~M.}, \bibinfo{author}{{a}o Filho, E. L.~F.} \&
  \bibinfo{author}{de~Ara\'{u}jo, C.~B.}
\newblock \bibinfo{title}{Characterization of topological charge and orbital
  angular momentum of shaped optical vortices}.
\newblock \emph{\bibinfo{journal}{Opt. Express}} \textbf{\bibinfo{volume}{22}},
  \bibinfo{pages}{30315--30324} (\bibinfo{year}{2014}).

\bibitem{Allen200067}
\bibinfo{author}{Allen, L.} \& \bibinfo{author}{Padgett, M.}
\newblock \bibinfo{title}{The {P}oynting vector in {L}aguerre-{G}aussian beams
  and the interpretation of their angular momentum density}.
\newblock \emph{\bibinfo{journal}{Opt. Commun.}}
  \textbf{\bibinfo{volume}{184}}, \bibinfo{pages}{67 -- 71}
  (\bibinfo{year}{2000}).

\bibitem{Bolda98}
\bibinfo{author}{Bolda, E.~L.} \& \bibinfo{author}{Walls, D.~F.}
\newblock \bibinfo{title}{Detection of vorticity in {B}ose-{E}instein condensed
  gases by matter-wave interference}.
\newblock \emph{\bibinfo{journal}{Phys. Rev. Lett.}}
  \textbf{\bibinfo{volume}{81}}, \bibinfo{pages}{5477--5480}
  (\bibinfo{year}{1998}).

\bibitem{Amo11}
\bibinfo{author}{Amo, A.} \emph{et~al.}
\newblock \bibinfo{title}{Polariton superfluids reveal quantum hydrodynamic
  solitons}.
\newblock \emph{\bibinfo{journal}{Science}} \textbf{\bibinfo{volume}{332}},
  \bibinfo{pages}{1167--1170} (\bibinfo{year}{2011}).

\bibitem{Hivet12}
\bibinfo{author}{Hivet, R.} \emph{et~al.}
\newblock \bibinfo{title}{Half-solitons in a polariton quantum fluid behave
  like magnetic monopoles}.
\newblock \emph{\bibinfo{journal}{Nature Phys.}} \textbf{\bibinfo{volume}{8}},
  \bibinfo{pages}{724--728} (\bibinfo{year}{2012}).

\bibitem{Boulier15}
\bibinfo{author}{Boulier, T.} \emph{et~al.}
\newblock \bibinfo{title}{Vortex chain in a resonantly pumped polariton
  superfluid}.
\newblock \emph{\bibinfo{journal}{Sci. Rep.}} \textbf{\bibinfo{volume}{5}}
  (\bibinfo{year}{2015}).

\bibitem{Cancellieri14}
\bibinfo{author}{Cancellieri, E.} \emph{et~al.}
\newblock \bibinfo{title}{Merging of vortices and antivortices in polariton
  superfluids}.
\newblock \emph{\bibinfo{journal}{Phys. Rev. B}} \textbf{\bibinfo{volume}{90}},
  \bibinfo{pages}{214518} (\bibinfo{year}{2014}).

\bibitem{hivet2014interaction}
\bibinfo{author}{Hivet, R.} \emph{et~al.}
\newblock \bibinfo{title}{Interaction-shaped vortex-antivortex lattices in
  polariton fluids}.
\newblock \emph{\bibinfo{journal}{Phys. Rev. B}} \textbf{\bibinfo{volume}{89}},
  \bibinfo{pages}{134501} (\bibinfo{year}{2014}).

\bibitem{dominici14}
\bibinfo{author}{Dominici, L.} \emph{et~al.}
\newblock \bibinfo{title}{Vortex and half-vortex stability in coherently driven
  spinor polariton fluid}.
\newblock \emph{\bibinfo{journal}{arXiv preprint arXiv:1403.0487}}
  (\bibinfo{year}{2014}).

\bibitem{tkachenko66}
\bibinfo{author}{Tkachenko, V.~K.}
\newblock \bibinfo{title}{Stability of vortex lattices}.
\newblock \emph{\bibinfo{journal}{Sov. Phys. JETP}}
  \textbf{\bibinfo{volume}{23}}, \bibinfo{pages}{1049} (\bibinfo{year}{1966}).

\bibitem{coddington2003}
\bibinfo{author}{Coddington, I.}, \bibinfo{author}{Engels, P.},
  \bibinfo{author}{Schweikhard, V.} \& \bibinfo{author}{Cornell, E.~A.}
\newblock \bibinfo{title}{Observation of tkachenko oscillations in rapidly
  rotating bose-einstein condensates}.
\newblock \emph{\bibinfo{journal}{Phys. Rev. Lett.}}
  \textbf{\bibinfo{volume}{91}}, \bibinfo{pages}{100402}
  (\bibinfo{year}{2003}).

\end{thebibliography}
\end{document}